\documentclass[aps,epsfig,twocolumn]{revtex4}

\usepackage{graphicx}
\usepackage{dcolumn}
\usepackage{bm}

\newcommand{\bq}{\begin{equation}}
\newcommand{\eq}{\end{equation}}
\newcommand{\bqa}{\begin{eqnarray}}
\newcommand{\eqa}{\end{eqnarray}}
\newcommand{\nn}{\nonumber \\}

\begin{document}
\draft
\title{
Amperean Pairing Instability in the U(1) Spin Liquid State with Fermi Surface
and Application to $\kappa-(BEDT-TTF)_2 Cu_2 (CN)_3$
}

\author{
Sung-Sik Lee$^{1}$,
Patrick A. Lee$^{1}$
and
T. Senthil$^{1,2}$
}
\address{
~$^{1}$ Department of Physics, Massachusetts Institute of Technology,\\
Cambridge, Massachusetts 02139, U.S.A.\\
~$^{2}$ Center for Condensed Matter Theory, Department of Physics,
Indian Institute of Science, Bangalore 560 012, India\\
}
\date{\today}

\date{\today}

\begin{abstract}
Recent experiments on the organic compound
$\kappa-(BEDT-TTF)_2 Cu_2 (CN)_3$
raise the possibility that the system may be
described as a quantum spin liquid.
Here we propose a pairing
state caused by the `Amperean' attractive
interaction between spinons on a Fermi surface
mediated by the U(1) gauge field.
We show that this state can explain many of
the observed low temperature phenomena
and discuss testable consequences.
\end{abstract}
\maketitle

\newpage

The organic compound $\kappa-(BEDT-TTF)_2 Cu_2 (CN)_3$ shows great
promise as the first candidate\cite{SHIMIZU1,KUROSAKI} which
realizes the spin liquid state in dimension greater than
one\cite{ANDERSON}. It is a quasi two dimensional material where
each plane forms a half filled triangular lattice. While it is an
insulator, there is no magnetic long range order and it behaves
like a metal as far as the spin dynamics is concerned. The uniform
spin susceptibility and the spin lattice relaxation rate $1/T_1T$
are finite in the zero temperature limit\cite{SHIMIZU1,KAWAMOTO}.
Recently a specific heat measurement was reported which
extrapolates to a linear $T$ coefficient $\gamma$\cite{NAKAZAWA}.
The $\gamma$ and spin susceptibility forms a Wilson ratio close to
unity. While an alternative interpretation in terms of Anderson
localization has been proposed, we find that the variable range
hopping fit to the resistivity\cite{KAWAMOTO}, when combined with
the density of states derived from $\gamma$, imply a localization
length of $0.9$ lattice spacing. Such a short localization
requires very strong disorder which make this interpretation
implausible.

Theoretically, the existence of the spin liquid state has been
suggested from the studies of the extended t-J
model\cite{MOTRUNICH} and the Hubbard model\cite{IMADA,LEE,KYUNG}
on the triangular lattice on the insulating side of the Mott
transition. In the spin liquid state, the low energy effective
theory becomes the U(1) gauge theory coupled with spinon which
forms a Fermi surface\cite{LEE}. The spinon carries only spin but
not charge and it contributes to the specific heat and thermal
conductivity even in the insulating state. The system of the
gapless spinon and the gauge field is a non-Fermi liquid
state\cite{PLEE92,POLCHINSKY} and exhibits singular temperature
dependence of specific heat coefficient, $\gamma \sim T^{-1/3}$.
However, recent specific heat measurement does not show this
singular behavior\cite{NAKAZAWA}. Moreover, there exists a kink in
the specific heat around $6 K$, which suggests that there is a
peak in the electronic specific heat if the phonon part is assumed
to be smooth. Around the same temperature the uniform
susceptibility also shows a sharp drop before it saturates to a
finite value in the zero temperature limit. These suggest that
there is a phase transition or a cross-over from the high
temperature phase which is described by the non-Fermi liquid state
to a low temperature Fermi liquid state. In the present paper we
propose that the low temperature phase may be understood as a
novel paired state of spinons that arises out of the U(1) spin
liquid state with spinon Fermi surface.

Inspired by Ampere's discovery that two wires carrying parallel
currents attract each other\cite{AMP}, we note that the
interaction is attractive when the two spinons have parallel
momenta. We therefore explore the possibility of pairing two
spinons on the same side of the Fermi surface. Within a simple
mean field treatment we demonstrate such a pairing instability of
the spinon Fermi surface. The resulting state has a number of
properties that are attractive for an explanation of the
experiments in $\kappa-(BEDT-TTF)_2 Cu_2 (CN)_3$. In particular
the pairing gaps out the gauge field so that the unpaired portion
of the Fermi surface gives a linear specific
heat at low temperature. We discuss various consequences of our
proposal that may be tested in future experiments.

Consider the system of spinon with Fermi surface
interacting with (non-compact) U(1) gauge field in 2+1D,
\bq
{\cal L}  =
  \psi_\sigma^* (  \partial_0 - i \phi - \mu ) \psi_\sigma
 + \frac{1}{2m} \psi_\sigma^* ( -i {\bf \nabla } -  {\bf a})^2 \psi_\sigma
  + \frac{1}{4g^2} f_{ij} f_{ij}  \nn
\label{eq:lag}
\eq
Here $x_0$ is the imaginary time and ${\bf
x}=(x_1, x_2)$ is the 2d spatial coordinate. $\psi_\sigma$ is the
spinon field with spin $\sigma$ and $\mu$, the chemical potential.
Repeated spin indices are summed. $a_i = (\phi, {\bf a})$ is the
U(1) gauge field with $i=0,1,2$, $f_{ij}$, the field strength
tensor and $g$, the gauge coupling. We expect $g^2$ to be
proportional to the charge gap and will ignore the last term in
Eq. (\ref{eq:lag}) in the following. We choose the Coulomb gauge
where ${\bf \nabla} \cdot {\bf a} = 0$. We are interested in the
stability of local Fermi surface in the momentum space and we
focus on a patch of Fermi surface which is centered at a momentum
${\bf Q}$ with $| {\bf Q} | = k_F$. Therefore we integrate out the
spinon fields except for those in the patch. The massless spinons
screen the temporal gauge field $\phi$. However, the transverse
gauge field is not screened and it can mediate a long range
interaction between spinons. The dressed propagator  of the
transverse gauge field is given by
\bq D(k)  =  \frac{1}{ \gamma_o
\frac{ |k_0| }{ \sqrt{ |{\bf k}|^2 + (k_0/{\bar v_F})^2 } +
|k_0|/{\bar v_F}} + \chi_d |{\bf k}|^2 }
\label{eq:gp}
\eq
where
$k=(k_0,{\bf k})$ is energy-momentum vector and $ \gamma_o =
\frac{{\bar v_F} {\bar  m}}{\pi} $ and $\chi_d = \frac{1}{12 \pi
{\bar m}}$ are the Landau damping and diamagnetic susceptibility
respectively. ${\bar v_F}$ is the Fermi velocity and ${\bar m}$,
the mass which are averaged over the Fermi surface which has been
integrated out. The transverse gauge field mediates an interaction
between spinons
\bqa S_{int} & = & -\frac{1}{2V\beta}
\sum_{p_1,p_2,k} D(k) \frac{ ( {\bf p}_1 \times \hat {\bf k} )
\cdot ( {\bf p}_2 \times \hat {\bf k} ) }{m^2}  \nn && \times
\psi_{\sigma p_1+k}^* \psi_{\sigma p_1} \psi_{\sigma^{'} p_2-k}^*
\psi_{\sigma^{'} p_2},
\eqa
where $V$ is the volume of the system,
$\beta = 1/(k_B T)$ and $m$, the mass of the spinon in the
vicinity of ${\bf Q}$ on the Fermi surface. It is noted that $m$
is generally different from $\bar m$ if the Fermi surface is not
perfectly spherical. Motivated by the Amperean attraction,
consider the pairing of two spinons with energy-momenta,
$p_1 = Q +  p$, $p_2 =  Q - p$,
where $2 Q$ is the net energy-momentum of the pair
with $Q_0 = 0$, $|{\bf Q}| = k_F$
and $p$, the relative energy-momentum with $| {\bf p} | <<  |{\bf Q}|$.
Note that the pair is made of two spinons on the same side of the Fermi surface and
it carries a large net momentum of $2k_F$.
We decompose the two body interaction into pairing channel by introducing
the Hubbard Stratonovich field $\Delta_{p}^{\sigma^{'} \sigma}$,
\bqa
S_{int} & = &
\frac{1}{2V\beta}
\sum_{p,p^{'}}
v(p^{'}-p)
\Bigl[
\Delta_{p^{'}}^{\sigma^{'} \sigma *}
\Delta_{p^{}}^{\sigma^{'} \sigma *} \nn
&&
- \Delta_{p^{'}}^{\sigma^{'} \sigma *}
\psi_{\sigma^{'} Q+p}
\psi_{\sigma^{} Q-p}
- c.c.
\Bigr],
\eqa
where $v(k) = \frac{ | {\bf Q} \times \hat {\bf k} |^2  }{m^2} D(k) $ and
we used $ ( {\bf Q} + {\bf p} ) \times \hat {\bf k}  \approx  {\bf Q} \times \hat {\bf k}$.
Pairing may occur in the singlet channel, i.e.
$\Delta_{p^{}}^{\uparrow \downarrow }  =  \Delta_{p}$,
$\Delta_{p^{}}^{\downarrow \uparrow }  =  -\Delta_{-p}$
and $\Delta_{p^{}}^{\uparrow \uparrow } = \Delta_{p^{}}^{\downarrow \downarrow } = 0$,
in which case $\Delta_p$ is an even function of ${\bf p}$,
or the triplet channel where $\Delta_p$ is odd in ${\bf p}$.
Now we integrate out the rest of the spinon field to obtain the Landau-Ginzburg free energy density,
\bqa
f[ \Delta ] & = & \Delta^\dagger ( v - v \Pi v) \Delta + O(\Delta^4).
\eqa
Here every product is a contraction of energy-momentum indices with a measure $\frac{1}{V \beta}$.
$\Delta$ is a vector with component $\Delta_p$,
and $v$, $\Pi$ are matrices with elements
$v_{p^{'},p} = v(p^{'}-p)$ and $\Pi_{p^{'},p} = V \beta g(Q+p) g(Q-p) \delta_{p^{'},p}$.
$g(p)$ is the spinon propagator given by
$ g(p) =  \frac{1}{ i \left( p_0 + \lambda  | p_0 |^{2/3} sgn( p_0 ) \right) + \epsilon_{\bf p} }$
with $\lambda = \frac{v_F}{2 \pi \sqrt{3} \chi_d^{2/3} \gamma_o^{1/3}}$
and $\epsilon_{\bf p} $, the spinon energy dispersion.
Here $v_F$ is the Fermi velocity at the patch which generally differs from the averaged one (${\bar v_F}$).

The system is unstable against developing pairing amplitude when
an eigenvalue of the kernel $( v - v \Pi v)$ becomes zero or negative.
Defining $\Phi = v \Delta$,
we write the eigenvalue equation
\bq
E_{pair} \Phi = v \Pi \Phi,
\label{eq:mt}
\eq
where $E_{pair}$ is the eigenvalue and $\Phi$ is the eigenvector.
Along the direction of the eigenvector,
the free energy density becomes
$f[ \Delta ]  \approx  \Delta^\dagger ( 1 - v \Pi ) \Phi =  ( 1 - E_{pair} ) \Delta^\dagger v \Delta$
and the system becomes unstable if $E_{pair}>1$.
The components of $\Phi$ has the unit of energy
and Eq. (\ref{eq:mt}) is nothing but the linearized self-consistent equation
for the anomalous self energy which gives rise to the gap
in the quasiparticle spectrum.
At zero temperature and in the thermodynamic limit, the matrix equation Eq. (\ref{eq:mt})
becomes an integral equation,
\bqa
E_{pair} \Phi(p) & = &
\int \frac{dp^{'}}{(2\pi)^3} v(p-p^{'}) g(Q+p^{'}) g(Q-p^{'}) \Phi( p^{'}). \nn
\label{eq:phi0}
\eqa
First, we approximately solve the equation analytically by
guessing an Ansatz for the eigenvector. Then we will check the
validity of the analytic solution by solving the equation
numerically without assuming a specific form of the eigenvector.
We first consider singlet pairing.

\begin{widetext}

\begin{figure}
        \includegraphics[height=9cm,width=10cm]{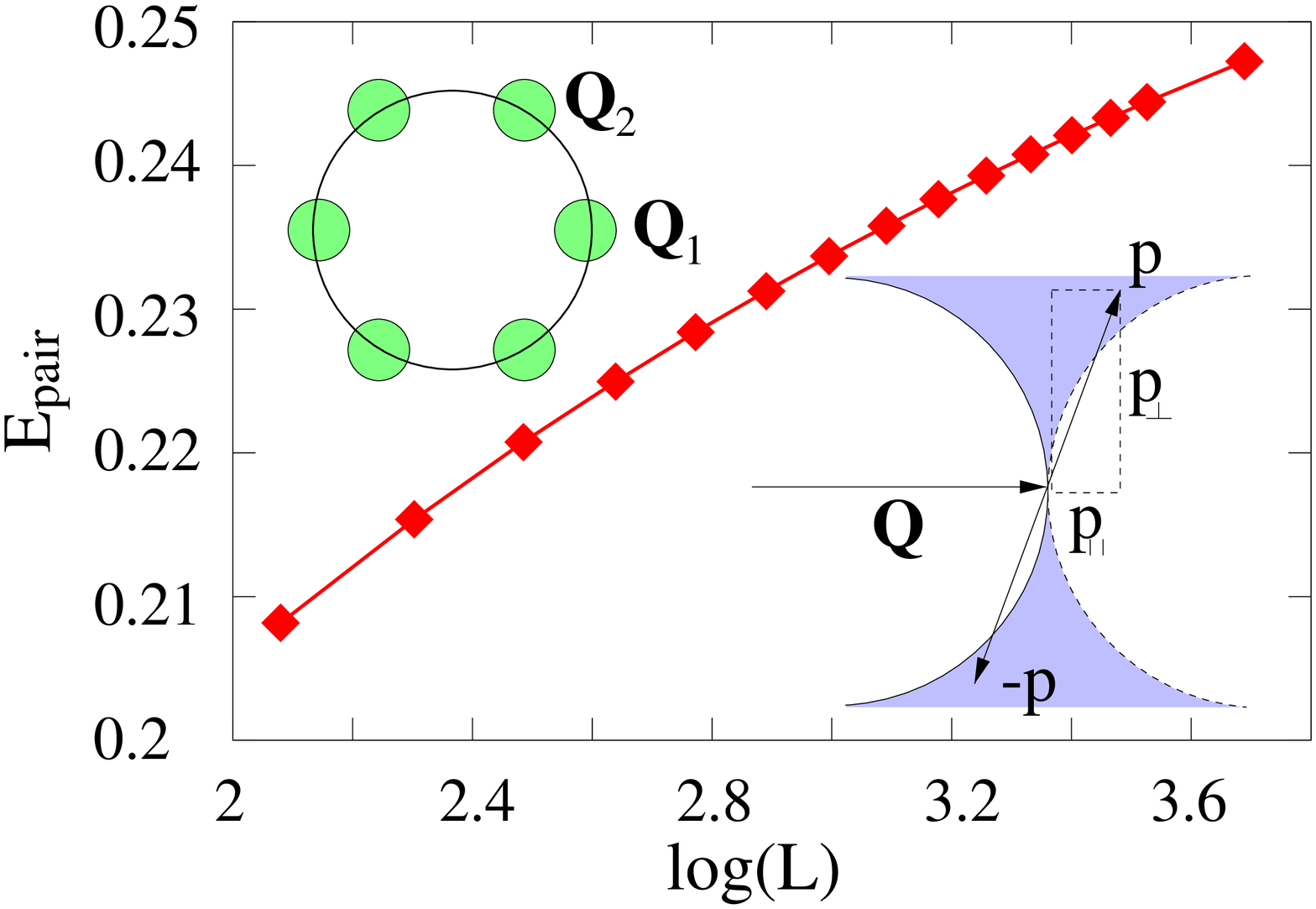}
\caption{
}
\label{fig:ep}
\end{figure}

\end{widetext}

In order to guess the form of the eigenvector, we determine the
important region of integration for ${\bf p^{'}}$ in Eq.
(\ref{eq:phi0}). If the pairing interaction were instantaneous and
the spinon did not have the frequency dependent self energy
correction, the $p^{'}_0$ integration would impose the constraint
that both of the constituent spinons of a pair should be on the
outside of the Fermi surface, that is, $| v_F p^{'}_\parallel |  <
\frac{ p^{'2}_\perp}{2m}$, where $p^{'}_\parallel$ ($p^{'}_\perp$)
is the momentum along (perpendicular to) the ${\bf Q}$ as is shown
in the bottom inset of Fig. \ref{fig:ep}. In the presence of the
frequency dependent interaction and spinon self energy, the sharp
constraint is smeared out. However, the dominant contribution of
the momentum integration still come from the region $| v_F
p^{'}_\parallel | < \frac{p^{'2}_\perp}{2m}$ which is denoted as
the shaded area in the bottom inset of Fig. \ref{fig:ep}. This has been
checked by performing a numerical integration of $p^{'}_0$.
Knowing the important region for ${\bf p^{'}}$ we consider an
approximate Ansatz, $\Phi(p_0, p_\perp, p_\parallel )  =  \tilde
\Phi( p_0, p_\perp ) \Theta( p_\perp^2/m - |v_F p_\parallel| )$,
where we take the range of $p_\parallel$ twice larger than $| v_F
p_\parallel | < \frac{p^{2}_\perp}{2m}$ in order to take into
account the smearing effect. This ansatz is singular at the curve
$p_\parallel = \frac{p_\perp^2}{m}$ which includes the point
$p_\parallel = p_\perp = 0$. A better treatment will smear out
this singularity. For this Ansatz, the typical momentum transfer
$k = p^{'} - p$ also satisfies the condition $|k_\parallel | <
\frac{k_\perp^2}{k_F} << k_\perp$ and we can ignore the
$k_\parallel$ dependence in the gauge propagator. We can also
replace $\frac{ | {\bf Q} \times \hat {\bf k} |^2 }{m^2}$ by
$v_F^2$ because ${\bf k}$ is almost perpendicular to ${\bf Q}$.
We perform the $k_\parallel$ integration in the Eq. (\ref{eq:phi0})
and we obtain
\bqa
&& E_{pair} \tilde \Phi(p_0, p_\perp)  =
\frac{v_F}{(2\pi)^3} \int dk_0 \int dk_{\perp} \frac{|k_\perp|}{
\gamma_o |k_0| + \chi_d | k_\perp|^3 }  \nn && \times \frac{m}{
(k_\perp + p_\perp)^2  }
\ln \left( 1 + \frac{  8 \left[
\frac{( k_\perp + p_\perp)^2}{2m}
\right]^2 } { \left[
\frac{( k_\perp + p_\perp)^2}{2m}
\right]^2   +   \lambda^2  | k_0 + p_0 |^{4/3}  } \right) \nn &&
\times \tilde \Phi(p_0+k_0, p_\perp+k_\perp).
\label{eq:phi3}
\eqa
Here we use the simpler form of gauge propagator which is obtained
from Eq. (\ref{eq:gp}) in the limit $v_F |{\bf k}| >> |k_0|$ and
we keep only the leading frequency dependent term in the spinon
propagator. Therefore, the integration for the energy/momentum
should be understood as having a ultraviolet cut-off of the order
of the Fermi energy/momentum.

It is noted that the right hand side of Eq. (\ref{eq:phi3}) is
smooth as a function of $p_0$ and depends on $p_0$ very weakly.
Therefore we ignore the $p_0$ dependence in the kernel and
consider a frequency independent eigenvector. The gauge propagator
$\frac{|k_\perp|}{ \gamma_o |k_0| + \chi_d | k_\perp|^3 }$ is
sharply peaked at $k_\perp \sim \left( \gamma_o|k_0| / \chi_d
\right)^{1/3}$ as a function of $k_\perp$ and can be approximated
by a delta function $\left( \gamma_o \chi_d^2 |k_0| \right)^{-1/3}
\sum_{s= \pm 1} \delta \left( k_\perp  - s ( \gamma_o |k_0| /
\chi_d )^{1/3} \right)$ and we can perform the $k_\perp$
integration. Changing the integration variable $k_0$ by $t = s
\left| \frac{\gamma_o}{\chi_d} k_0 \right|^{1/3}$ we obtain the
eigenvalue equation
\bqa && E_{pair} \tilde \Phi(p_\perp)  = 6
\frac{m v_F  }{  (2 \pi)^3 \gamma_o } \int d t \frac{|t|}{ |  t +
p_\perp |^2  }  \nn && \times \ln \left( 1 + \frac{  8 |  t +
p_\perp |^4    } { |  t + p_\perp |^4   +   A  |t|^4  } \right)
\tilde \Phi ( t + p_\perp ), \label{eq:phi52} \eqa where $A =
\left[ 2 m \lambda (\chi_d/\gamma_o)^{2/3} \right]^2$ is a
dimensionless constant. If we consider the spinon pair right on
the Fermi surface ($p_\perp = 0$) and use the Ansatz
$\Phi(p_\perp) = const.$, the right hand side of Eq.
(\ref{eq:phi52}) is logarithmically divergent. This signifies that
we can find an eigenvector which has an arbitrarily large
eigenvalue. However, the momentum independent Ansatz cannot
satisfy the eigenvalue equation because the kernel strongly
depends on $p_\perp$. In view of the singular dependence of the
kernel on $p_\perp$, we consider an Ansatz
\bq \tilde \Phi(
p_\perp) = \tilde \Phi_0 \frac{1}{ |p_\perp |^\alpha},
\label{eq:an} \eq where $\alpha$ should be smaller than $3/2$ in
order for the eigenvector to be normalizable. This Ansatz solves
the eigenvalue equation with the eigenvalue $E_{pair} \sim
\frac{6}{(2 \pi)^2 c} \ln \left( 1 + \frac{24 c^2}{ 3c^2 + 1}
\right) \frac{ 1 }{  \alpha }$, where $c = \frac{ {\bar m} {\bar
v_F}}{ m v_F}$ measures the local curvature of the Fermi surface.
For small enough $\alpha$ the eigenvalue can be arbitrarily large.
Thus within the present mean field treatment there is a pairing
instability.

Now we check the validity of the analytic solution by solving the eigenvalue equation numerically.
We do not assume a specific form of $\Phi(p)$ in Eq. (\ref{eq:phi0}).
Then the natural cut off for the $k_\parallel = p_\parallel-p_\parallel^{'}$
integration is $k_\perp$ not $k_\perp^2/m$ because
the coupling $\frac{ | {\bf Q} \times \hat {\bf k} |^2  }{m^2}$
becomes small for $k_\parallel > k_\perp$.
Ignoring the $k_\parallel$ dependence of the gauge propagator,
we can cast the equation into a 2D integral equation by applying
$\int_{-p_\perp}^{p_\perp} \frac{d p_\parallel}{2\pi} g(Q+p) g(Q-p)$
on both sides of the Eq. (\ref{eq:phi0}).
The resulting equation involves only two integrations and one can easily diagonalize
the kernel $M(p,p^{'})$ numerically to find the eigenvalue
and the eigenvector.
The largest eigenvalue corresponds to
singlet pairing (even $\Delta_p$)
and as shown in Fig. \ref{fig:ep}
increases logarithmically with increasing $L$,
where $L$ determines the mesh of the discrete energy and momentum
as $\Delta k = 2 \pi / L$.
The eigenvalue will become larger than $1$ for a large enough $L$
and there exists pairing instability in the thermodynamic limit.
The infrared divergence of the eigenvalue in the thermodynamic limit is consistent
with the analytic result that the eigenvalue diverges as $\alpha \rightarrow 0$.
Although not shown here,
the numerically calculated eigenvector is qualitatively consistent with the analytic Ansatz
with $\alpha < 1$.
The second largest eigenvalue corresponds to triplet pairing and is also
logarithmically divergent with a slope $10$ times smaller than that shown
in Fig. \ref{fig:ep}.
In the rest of the paper we assume singlet pairing, even though
we should be mindful that triplet pairing is also unstable and may be
preferred by short range repulsion.

The origin of the mean field pairing instability should be
contrasted with conventional superconductors where electrons with
momenta ${\bf p}$ and -${\bf p}$ form a pair which uses the whole
Fermi surface to lower its energy.
In the present case, spinon pairs carry a momentum of the order of $2
k_F$. While the LOFF state also carries finite
momentum\cite{LOFF}, our case is fundamentally different because
pairing fermions on the same side of the Fermi surface severely
restricts the available phase space. Consequently, the pairing
instability found here can happen only if the pairing interaction
is sufficiently singular.

Now consider possible instabilities in the particle-hole channel.
In the leading order of perturbation, the $2k_F$ vertex function
is logarithmically divergent if both the external momenta are on
the Fermi surface and energy transfer is zero\cite{ALTSHULER}. The
singular vertex function enhances the susceptibility of the spin
density wave with the momentum $2k_F$.  A mean field treatment
similar to that above reveals the existence of $2k_F$ density wave
instabilities in both the singlet and triplet channels, albeit
with non-trivial momentum dependence for the internal wavefunction
for the particle-hole pair.
We emphasize that the mean field
instability occurs in a channel where the eigenvector has a
specific momentum dependence. The system can remain stable in
other channel.
We point out that the $2 k_F$ instability preserves
the U(1) gauge structure.
Furthermore the Amperean pairing is favored for large local curvature
$c$ while the $2k_F$ instability prefers small $c$.

Theoretically the mean field results above should be regarded as
merely suggestive of possible low temperature instabilities of the
spinon Fermi surface state. Lacking a better theoretical treatment
we will take experiments as a guide for further discussion.

The NMR measurement does not observe the line broadening expected
for the incommensurate spin density wave\cite{SHIMIZU1,SHIMIZU2}.
Thus we discard the triplet spinon density wave. The singlet
spinon density wave state will have a non-Fermi liquid specific
heat unless a further Amperean pairing instability develops at low
temperature to restore Fermi liquid behavior (see below). However
in the latter case two separate transitions would have been
expected as a function of temperature (for instance as visible
signatures in the specific heat) which is not observed. Therefore,
we focus on the scenario where only the spinon pairing occurs and
explore some consequences.

First, in the paired state gaps will open on the patches of the
Fermi surface where the pairing occurs.
The momentum point at which the pairing instability first occurs
depends on the details of the Fermi surface.
In general there will be a number of preferred
points related by hexagonal symmetry. Once the pairing occurs on
parts of the Fermi surface, the U(1) gauge group is reduced to
$Z_2$. Since the $Z_2$ gauge field is gapped, the low energy
theory becomes the Fermi liquid theory and the remaining Fermi
surface can remain gapless without further instability. This is
consistent with the observation that there exists a finite
specific heat coefficient $\gamma$ in the zero temperature limit
rather than the singular $T^{-1/3}$ behavior. The proposed spinon
pair state will generically break lattice translation, rotational
or even time reversal symmetries.
As shown in the top inset of Fig. \ref{fig:ep}, suppose the pairing
occurs at two distinct favored momenta ${\bf Q}_1$ and ${\bf Q}_2$ with
$\Delta_1$, and $\Delta_2$ the corresponding pairing order
parameters. These are of course not gauge invariant but the
gauge-invariant combination $(\Delta_1)^* (\Delta_2)$ is at
non-zero momentum $2 ( {\bf Q}_2 - {\bf Q}_1 )$
and is also condensed as $\Delta_1$
and $\Delta_2$ are individually condensed.
This represents a spontaneous breakdown of lattice symmetry.
As we are discussing a spin system,
this order corresponds to an incommensurate version of
the valence bond solid (spin Peierls state).
However we emphasize that this broken lattice symmetry state
coexists with fractionalized spinons.
The broken lattice symmetry
implies a finite temperature phase transition.
However due to the
incommensurate ordering the transition should be 2d X-Y like and
shows no observable singularity in the specific heat.
The translational symmetry breaking should couple to
lattice distortion and may be observable by X-ray scattering.

A key prediction is that the low temperature thermal conductivity
$\kappa \sim T$ like in a metal in contrast to the vanishing
thermal conductivity expected in an Anderson insulator or the
enhanced $\kappa \sim T^{\frac{1}{3}}$ for the spinon Fermi
surface state with a gapless $U(1)$ gauge field\cite{LEE}.

One may think that the sharp drop of the uniform spin
susceptibility below $6K$\cite{SHIMIZU1} can be explained from the
reduction of the spinon density of state caused by spinon pairing.
However, this explanation is not correct for the following reason.
In contrast with BCS theory but in common with the LOFF state, the
Amperean pairing is not destroyed by the Zeeman limiting field
because the spinon with up spin with momentum $|{\bf Q}_\uparrow|
= |{\bf Q}| + \mu_B H / v_F$ and the spinon with down spin with
momentum $|{\bf Q}_\downarrow| = |{\bf Q}| - \mu_B H / v_F$ can
both be on the Fermi surface and paired without the energy cost of
the Zeeman energy. This property is crucial in explaining the lack
of field dependence up to $8T$ in the specific
heat\cite{NAKAZAWA}. However, it follows that the uniform
susceptibility should not be affected by the opening of the
pairing gap. We suggest that the reduction of the susceptibility
may come from a contribution of the gauge field which is gapped
out at low temperature. Details will be discussed
elsewhere\cite{CODY}. The spinon pairing state is not a
superconductor because the spinon does not carry charge. However,
if the charge gap is suppressed by driving the system across the
Mott transition point with pressure\cite{KUROSAKI}, Bose
condensation of the charge degrees of freedom converts the
Amperean pairing state to a real superconductor. One signature of
this unconventional superconductor is that the Knight shift will
hardly change across the transition temperature, which is highly
unusual for singlet pairing. This signature is consistent with
recent data\cite{KNIGHT}.

In conclusion, aside from intrinsic theoretical interest, our
proposal of a novel spin liquid state with paired spinons explains
many of the unusual low temperature behaviors in
$\kappa-(BEDT-TTF)_2 Cu_2 (CN)_3$ and is amenable to further
experimental tests.

PAL acknowledge the support by NSF DMR-0517222. TS acknowledges
support from a DAE-SRC Outstanding Investigator Award in India,
the Alfred P. Sloan Foundation, and The Research Corporation. We
thank M. P. A. Fisher, L. B. Ioffe, Y. B. Kim for helpful
discussions, and K. Kanoda and Y. Nakazawa for sharing their data
prior to publication.


\newpage
\centerline{FIGURE CAPTIONS}
\begin{itemize}

\item[Fig. 1]
(color online)
The largest eigenvalue of $M(p,p^{'})$ as a function of
the system size $L$ where the mesh of the discrete energy and momentum
is given by $\Delta k = 2 \pi / L$.
Top inset : schematic picture of
partial gapping of the spinon Fermi surface.
Bottom inset :
definition of $p_\parallel$ and $p_\perp$.

\end{itemize}

\newpage

\begin{thebibliography}{27}

\bibitem{SHIMIZU1} Y. Shimizu, K. Miyagawa, K. Kanoda, M. Maesato and G. Saito, Phys. Rev. Lett. {\bf 91}, 107001 (2003).
\bibitem{KUROSAKI} Y. Kurosaki, Y. Shimizu, K. Miyagawa, K. Kanoda and G. Saito, Phys. Rev. Lett. {\bf 95}, 177001 (2005).
\bibitem{ANDERSON} P. W. Anderson, Science {\bf 235}, 1196 (1987); P. Fazekas and P. W. Anderson, Philos. Mag. {\bf 30}, 432 (1974).
\bibitem{KAWAMOTO} A. Kawamoto, Y. Honma and K. I. Kumagai, Phys. Rev. B {\bf 70}, 060510(R) (2004).
\bibitem{NAKAZAWA} Y. Nakazawa et.al., unpublished
\bibitem{MOTRUNICH} O. I. Motrunich, Phys. Rev. B {\bf 72}, 045105 (2005).
\bibitem{IMADA} H. Morita, S. Watanabe and M. Imada, J. Phys. Soc. Jpn. {\bf 71}, 2109 (2002).
\bibitem{LEE} S.-S. Lee and P. A. Lee, Phys. Rev. Lett. {\bf 95}, 036403 (2005).
\bibitem{KYUNG} B. Kyung and A.-M. S. Tremblay, cond-mat/0604377.
\bibitem{PLEE92} P. A. Lee and N. Nagaosa, Phys. Rev. B {\bf 46} 5621 (1992).
\bibitem{POLCHINSKY} J. Polchinski, Nucl. Phys. B {\bf 422}, 617 (1994).
\bibitem{AMP} A.-M. Ampere, Annales de chimie et de physique {\bf 15}, 59 (1820); ibid. {\bf 15}, 170 (1820).
\bibitem{LOFF} P. Fulde and R. A. Ferrell, Phys. Rev. {\bf 135}, A550 (1964); A. I. Larkin and Y. N. Ovchinnikov, Zh. Eskp. Teor. Fiz. {\bf 47}, 1136 (1964).
\bibitem{ALTSHULER} B. L. Altshuler, L. B. Ioffe and A. J. Millis, Phys. Rev. B {\bf 50}, 14048 (1994).
\bibitem{SHIMIZU2} Y. Shimizu, K. Miyagawa, K. Kanoda, M. Maesato and G. Saito, Phys. Rev. B {\bf 73}, 140407(R) (2006).
\bibitem{CODY} C. P. Nave, S.-S. Lee and P. A. Lee, in preparation.
\bibitem{KNIGHT} Y. Shimizu, H. Kasahara, K. Miyagawa, K. Kanoda, M. Maesato and G. Saito, preprint.

\end{thebibliography}
\end{document}